\documentclass[reprint,amsmath,amssymb,aps,prb,superscriptaddress]{revtex4-1}
\usepackage{units}
\usepackage{amsmath}
\usepackage{url}
\usepackage{comment}
\usepackage{natbib}
\usepackage{textcase}
\usepackage{amssymb}
\usepackage{graphicx}
\usepackage{bm}
\usepackage{float}
\usepackage{multirow,microtype,color,relsize,ulem}
\usepackage{color}
\usepackage{braket}
\usepackage[large]{subfigure}
\usepackage{hyperref}
\usepackage[utf8]{inputenc}
\usepackage[english]{babel}
\hypersetup{colorlinks=true,linkcolor=blue,citecolor=blue}
\hypersetup{linktocpage}

\begin{document}
	\title{Transport and spectral features in non-Hermitian open systems}
	
	\author{A. F. Tzortzakakis}
	\affiliation{Physics Department, University of Crete, 71003 Heraklion, Greece}
	\author{K. G. Makris}
	\affiliation{Physics Department, University of Crete, 71003 Heraklion, Greece}
	\affiliation{Institute of Electronic Structure and Laser, FORTH, 71110 Heraklion, Crete, Greece}
	\author{A. Szameit}
	\affiliation{Institut f{\"u}r Physik, Universit{\"a}t Rostock, 18059 Rostock, Germany}
	\author{E. N. Economou}
	\affiliation{Physics Department, University of Crete, 71003 Heraklion, Greece}
	\affiliation{Institute of Electronic Structure and Laser, FORTH, 71110 Heraklion, Crete, Greece}

\begin{abstract}
We study the transport and spectral properties of a non-Hermitian  one-dimensional disordered lattice, the diagonal matrix elements of which are random complex variables taking both positive (loss) and negative (gain) imaginary values: Their distribution is either the usual rectangular one or a binary pair-correlated one possessing, in its Hermitian version, delocalized states and unusual transport properties. Contrary to the Hermitian case, all states in our non-Hermitian system are localized. In addition, the eigenvalue spectrum, for the binary pair-correlated case, exhibits an unexpected intricate fractallike structure on the complex plane and with increasing non-Hermitian disorder, the eigenvalues tend to coalesce in particular small areas of the complex plane, a feature termed ``eigenvalue condensation". Despite the strong Anderson localization of all eigenstates, the system appears  to exhibit  transport not by diffusion but by a new mechanism  through sudden jumps between states located even at distant sites. This seems to be a general feature of open non-Hermitian random systems. The relation of our findings to recent experimental results is also discussed.
\end{abstract}

\date{\today}
\maketitle

\section{Introduction}

\par Anderson localization raising the possibility of suppression of diffusion in disordered media \cite{loc3}, is a fundamental phenomenon of wave physics, and has been extensively studied in both quantum and classical domain \cite{loc1,loc11,loc12,loc13,loc14,loc4,loc5,loc6}. Its importance in various fields, such as condensed matter physics, disordered photonics and imaging, Bose-Einstein condensates and acoustic waves is evident. However, with the exception of the random laser community \cite{rl1,rl2,rl3}, the majority of the studies regarding wave localization has been devoted to conservative systems, in which Hermiticity of the Hamiltonian is ensured. Whereas accurate control of the openness in many fields of wave physics is difficult or even impossible, photonics provides an ideal area where such control is possible by today's available experimental techniques.
\par In particular, the recent introduction of the concepts of parity-time ($\mathcal{PT}$) symmetry \cite{Bender1,Bender2,Bender3,PT1,PT2,PT3,PT4,PT5} and exceptional points \cite{EP1,EP2,EP3,EP4,EP5} in optics, which relies on the complex values of the index of refraction, has led to the development of a new research field, that of non-Hermitian photonics \cite{review1,review2,Pile,Gbur,review3,Longhi,review4,review5}. 
In particular, the openness of these systems can be described in terms of gain (amplification based on laser materials) and/or loss (intrinsic decay mechanism) and their delicate interplay leads to unexpected novel features.
The rich behavior of these structures has triggered a plethora of experimental realizations of various optical devices \cite{PT6,PT7,PT8,PT9,PT10,PT11,PT12,PT13,PT14,PT15,PT16,lin_unidirectional_2011, horsley_spatial_2015,konotop_families_2014,zhu_one-way_2013}.

\par Quite recently there has been a renewed interest for non-Hermitian Anderson localization problems \cite{Andreas,nhloc1,nhloc2,nhloc3,nhloc4,nhloc5}, since it was realized that in the context of optical physics one can experimentally realize linear random non-Hermitian Hamiltonians, away from the highly nonlinear regime of random lasers and the majority of abstract non-Hermitian random matrices. The proposed complex random discrete models can be considered the most relevant non-Hermitian analog of the Anderson original problem. In this case, the non-Hermiticity is a direct consequence of the complex nature of the index of refraction, whereas the coupling between nearest neighbors is real and fixed. Thus the fundamental questions of whether the eigenmodes are localized or not and  whether transport is possible, still remain open. Interestingly, in a recent novel experiment \cite{Alex} it was demonstrated that the non-Hermiticity of a random medium with a rectangular distribution of disorder can unexpectedly result to jumpy evolution dynamics despite the strong localization of all corresponding eigenfunctions. We emphasize that quite different non-Hermitian Anderson models have been previously examined in a number of related theoretical works \cite{nhloc6,nhloc7,nhloc8,nhloc9} (see Appendix A).

\section{Formulation}

\par
\begin{figure}[tb!]
	\centering
	\includegraphics[width=0.45\textwidth]{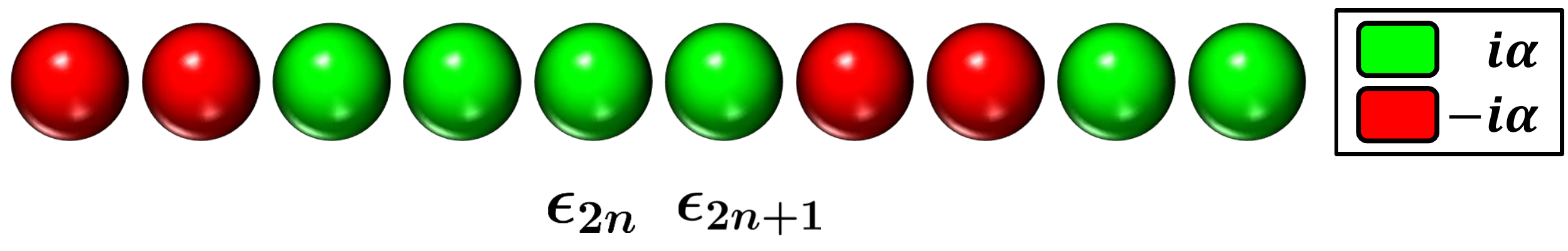}
	\caption{Schematic representation of the complex diagonal matrix elements for the case of the non-Hermitian random binary pair-correlated model of Eq.~(\ref{2b}).}
	\label{schem}
\end{figure}

 In this work we study for the first time the spectral and dynamic properties of one-dimensional waveguide lattices, which are characterized by non-Hermitian  disorder in the diagonal matrix elements {$\epsilon_{n}=\epsilon_{R,n}+i\epsilon_{I,n}$}; the imaginary part, $\epsilon_{I,n}$, of the latter has either a rectangular distribution, defined in Eq.~(\ref{2a}), usually centered around zero with a total width equal to $2W$ (being thus a direct non-Hermitian extension of Anderson original paper) or has a binary pair-correlated distribution, defined in Eq.~(\ref{2b}) and shown schematically  in Fig.~\ref{schem}, being the non-Hermitian extension of the extensively studied, random pair-correlated (dimer) model\cite{dim1,dim2,dim3,dim4}, which is the simplest disorder system that, in spite of being one-dimensional, still permits wavepacket delocalization, as it has also been demonstrated experimentally \cite{dim4} (see Appendix B and in particular Fig.~\ref{b1}). In our non-Hermitian binary pair correlated model binary disorder though, we show that delocalization is impossible, for any value of the complex diagonal elements. Furthermore, we find that such binary disordered system exhibits various unexpected features, such as fractallike spectrum,  as well as regions in the complex plane where many eigenvalue come arbitrarily close and form ``eigenvalue condensates". However, despite the strong Anderson localization exhibited for all distributions  employed in the present work for the random variables $\epsilon_{I,n}$,   a new kind of non-Hermitian transport dynamics seems to occur with sudden jumps between eigenstates localized even around distant sites. Our study may pave the way for future optical experiments that demonstrate the counter-intuitive dynamic properties of non-Hermitian disordered lattices.

\par The paraxial wave propagation in a 1D waveguide array of $N$ coupled channels, is described by the normalized coupled mode equations:
\begin{equation}
i\frac{\partial \psi_{n}}{\partial z}+c(\psi_{n+1}+\psi_{n-1})+\epsilon_{n}\psi_{n}=0
\label{par}
\end{equation}
where $z$ is the propagation distance, $\psi_{n}$ and $\epsilon_{n}$ are the envelope of the electric field and the propagation constant of the $n^{th}$ channel (which here plays the role of the on-site energy) and $c$ is the coupling constant between nearest neighbors, which as usually we have set $c=1$ for simplicity. For the spectral properties of the model, we will consider stationary right eigenstates of the form: $\psi_{n,j}(z)=u_{n,j}e^{i\omega_{j} z}$, where $\omega_{j}$ is a complex eigenvalue of the system. For our system the left eigenstates are complex conjugate of the right as is explained in more detail in Appendix A.
\par As mentioned,  the Hermitian version of our model (where all the $\epsilon_{n}$ are real random numbers), is either the original Anderson model or the  well known case of binary pair-correlated disorder possessing extended eigenstates (in spite of being one-dimensional) and exhibiting superdiffusive transport, as has been analytically and experimentally \cite{dim1,dim2,dim3,dim4} shown (see Appendix B).  In this work we set initially $\epsilon_{R,n}=\textnormal{Re}(\epsilon_{n})=0$ and choose $\epsilon_{I,n}=\textnormal{Im}(\epsilon_{n})$ to be a random variable with the following  distributions among others:
\begin{subequations}
	\begin{equation}
	\textnormal{p}(\epsilon_{I,n})=\begin{cases} \hspace{0.1cm}
	1/2W \quad \textnormal{if} \quad \epsilon_{I,n}\in(-W,W)\\
	\hspace{0.3cm}
	0 \quad \textnormal{otherwise}
	\end{cases} 
	\label{2a}
\end{equation}
or
	\begin{equation}
	\epsilon_{I,2n}=\epsilon_{I,2n+1}=\begin{cases} \hspace{0.3cm} i\alpha, \quad \textnormal{with} \quad \textnormal{p}_{1}=\frac{1}{2} \\
	-i\alpha, \quad \textnormal{with} \quad \textnormal{p}_{2}=\frac{1}{2}                 
	\end{cases}
	\label{2b}
	\end{equation}
\end{subequations}
with p$_{1,2}$ being the associated probabilities, and $\alpha$ a real, positive number describing the disorder's amplitude. Thus the distribution of Eq.~(\ref{2b}) is a binary one with short range  pair-correlation such that each pair of two consecutive sites have randomly either the value $i\alpha$ or $-i\alpha$ as shown schematically in Fig.~\ref{schem}. Other distributions similar to Eq.~(\ref{2a}) were also examined (see Appendix D).

\section{SPECTRAL  PROPERTIES FOR THE DISTRIBUTION SHOWN IN  Eq.~(\ref{2b})}

\par
\begin{figure}[tb!]
	\centering
	\includegraphics[clip,width=1\linewidth]{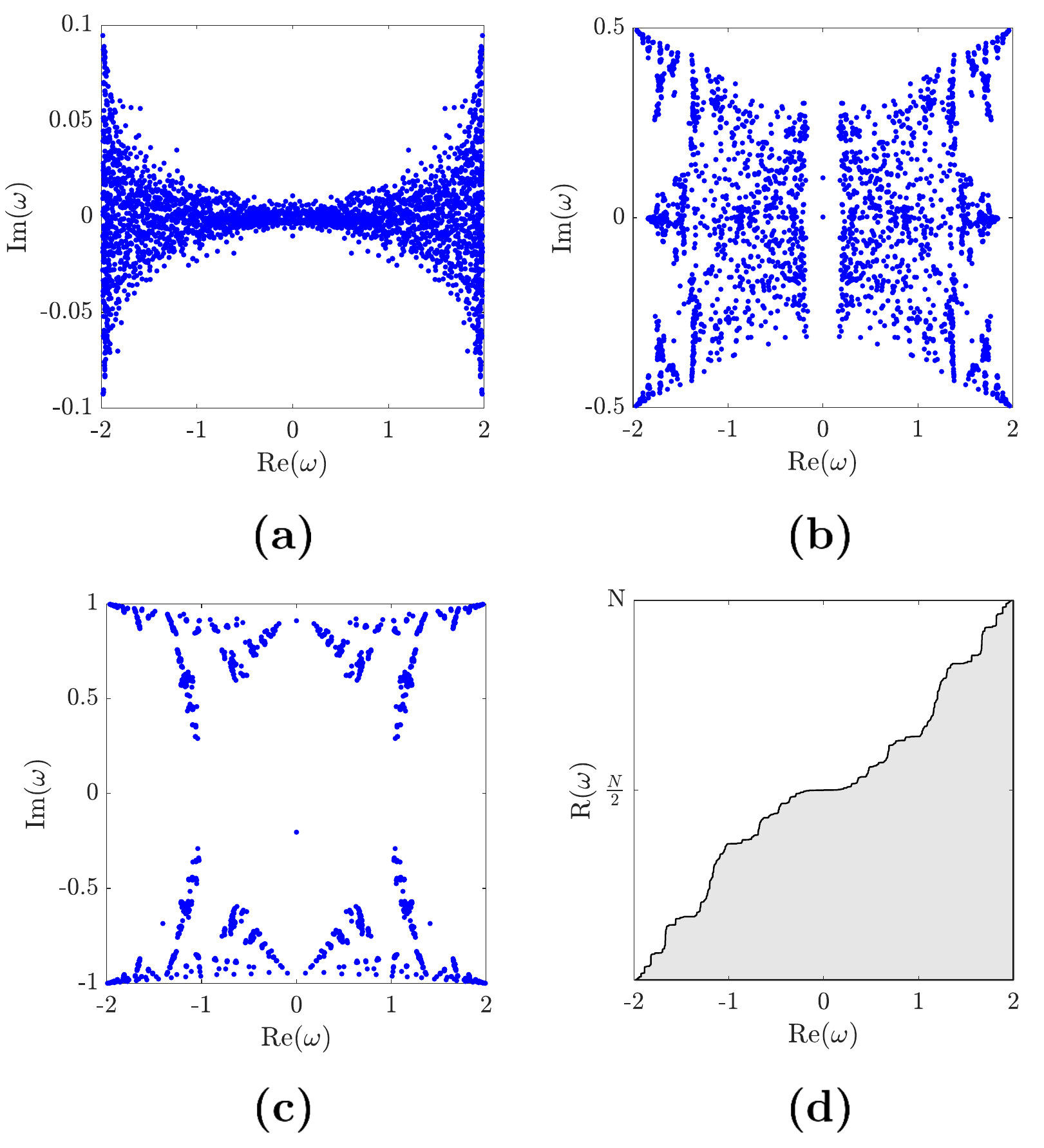}
	\caption{(a)-(c) Eigenvalue spectrum on the complex plane for the non-Hermitian random binary pair-correlated model and for (a) $\alpha=0.1$, (b) $\alpha=0.5$, and (c) $\alpha=1$. (d) Integral density of states R$(\omega)$ as a function of the real part Re$(\omega)$ of the eigenvalues [R$(\omega)$ counts all states with Re$(\omega_j)<\omega$]. All these results correspond to $N=3000$.}
	\label{spectr}
\end{figure}

 In Fig.~\ref{spectr} we present our result for the spectrum in the complex plane, as well as the integral density of states for various values of the randomness parameter $\alpha$. The spectrum for small values of $\alpha$ is concentrated near the real axis, except for the edges whose imaginary part extends through $(-\alpha,\alpha)$ [Fig.~\ref{spectr}(a)]. These results are reasonable, and can be intuitively associated with the density of states of the corresponding Hermitian problem.

\par However, for $\alpha=0.5$ the picture is quite different. The whole spectrum now tends to move away from the real axis and to form an intricate fractallike structure in the complex plane [Fig.~\ref{spectr}(b)], which resembles the spectrum of the quasiperiodic Harper model \cite{harp}, in exhibiting a similar regularity in spite of its randomness. In addition, a gap opens around the imaginary axis Re$(\omega)=0$. We note that it is not quite uncommon for non-Hermitian random matrices to exhibit intricate fractallike spectra in the complex plane\cite{nhloc9}. It must be pointed out that these features are associated with the binary pair-correlated character of our model; they disappear in the absence of pair-correlation and in the case of a rectangular distribution of the random variable. By increasing the value of $\alpha$ the eigenvalues do not extend over the whole complex plane, as one might expect, but they rather tend to ``collapse" into specific points of the complex plane [Fig.~\ref{spectr}(c)] leading to rather sparse spectrum. We term this behavior as ``eigenvalue condensation". Such eigenvalue coalescence can be directly shown if one plots the integral  density of states R$(\omega)$ as a function of the real part of the eigenvalues [Fig.~\ref{spectr}(d)], where $\textnormal{R}(\omega)$ counts all states with the real part of the corresponding eigenvalue less than $\omega$. We can clearly see the step like behavior of the eigenvalues due to their condensation. If we further increase the value of $\alpha$, then the eigenvalues gradually approach the lines with $\textnormal{Im}(\omega)=\pm\alpha$, i.e., they tend to take the form of the diagonal matrix elements as expected, and the intriguing yet unexplained pattern of the spectrum is lost.
\par
\begin{figure}[tb!]
	\centering
	\includegraphics[clip,width=1\linewidth]{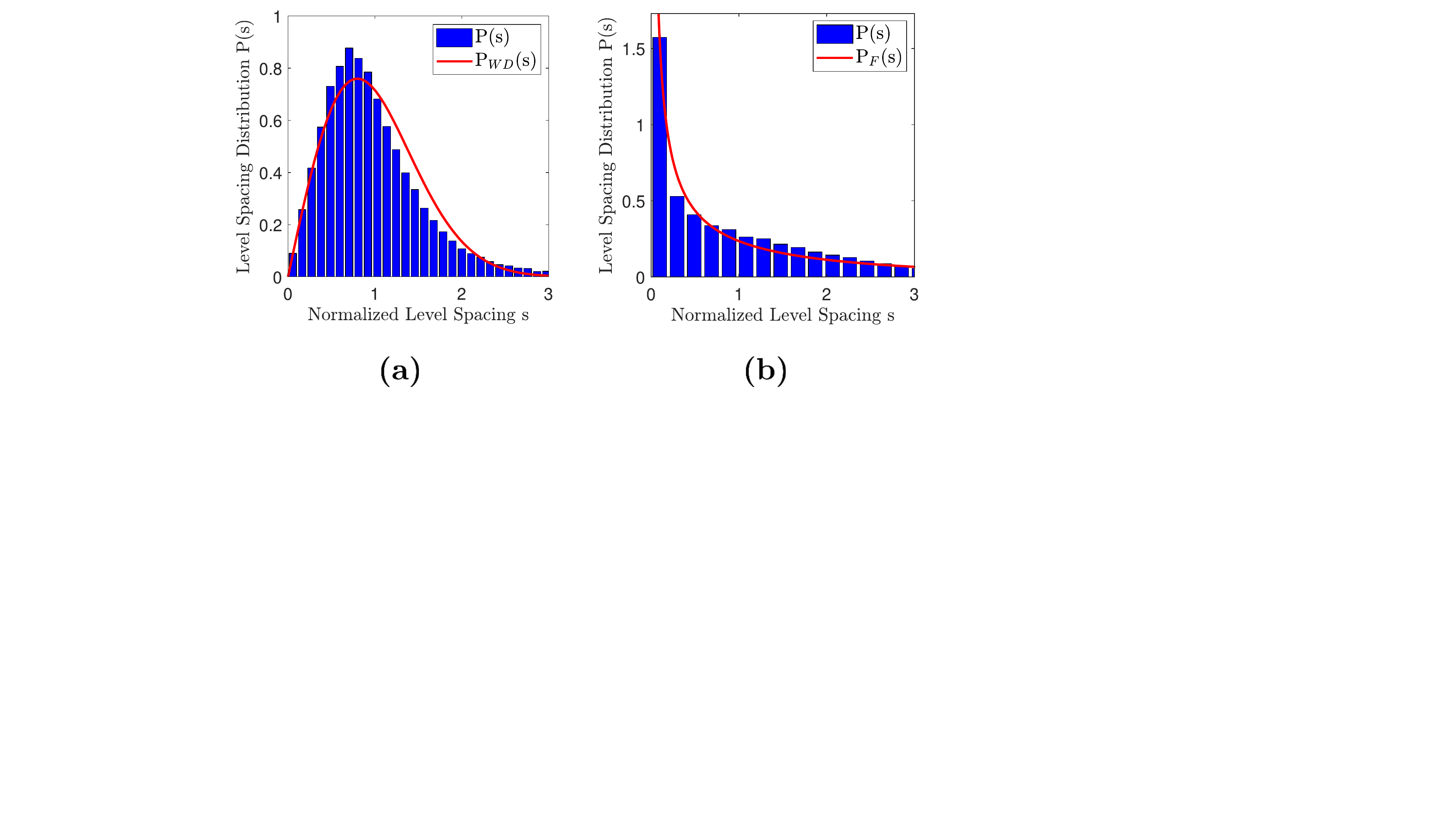}
	\caption{Normalized level spacing distribution P$(s)$, averaged over the whole spectrum, for (a) $\alpha=0.1$ compared to a Wigner-Dyson distribution (red curve, see App. A), and (b) $\alpha=0.5$, fitted with a function of the form of Eq.~(\ref{pf}) (red curve), with: $A\approx0.3$, $\beta \approx 0.72$ and $\lambda \approx 0.24$. An averaging over 50 realizations of disordered systems with $N=2500$ has been performed for these results.}
	\label{lvlsp}
\end{figure}

 In order to quantify the aforementioned change in the spectrum's picture (Fig.~\ref{spectr}), taking place for $\alpha$ between $0.1$ and $0.5$, we examine the level spacing distribution P$(s)$, where $s$ is the normalized minimum distance between two eigenvalues in the complex plane: $s|_{\omega_j}\equiv \min|\omega_{j}-\omega_{j'}|$ (Fig.~\ref{lvlsp}).
\par For $\alpha=0.1$ we get the expected Wigner-Dyson distribution: P$_{WD}(s)=\frac{\pi s}{2}e^{-\frac{\pi s^2}{4}}$ and the eigenvalues show level repulsion. On the other hand, if we set $\alpha=0.5$ the level spacing distribution resembles the Poisson distribution, P$_{P}(s)=e^{-s}$. However, in contrast to P$_{P}(s)$, our histogram [Fig.~\ref{lvlsp}(b)] seems to exhibit a possible singularity as $s\to0$. We found that the following expression:
\begin{equation}
	\textnormal{P}_F(s)= A s^{-\beta}e^{-\lambda s}~,
	\label{pf}
\end{equation}
provides a reasonable fitting to our data. We think that this behavior is not inconsistent with the aforementioned eigenvalue condensation, since a large number of eigenvalues tend to coalesce. We found that, by increasing the size $N$ of our system, the histograms tend to be closer to the corresponding analytic formulas.

\section{LOCALIZATION IN  OUR SYSTEMS}

\par
\begin{figure}[htb!]
	\centering
	\includegraphics[clip,width=0.8\linewidth]{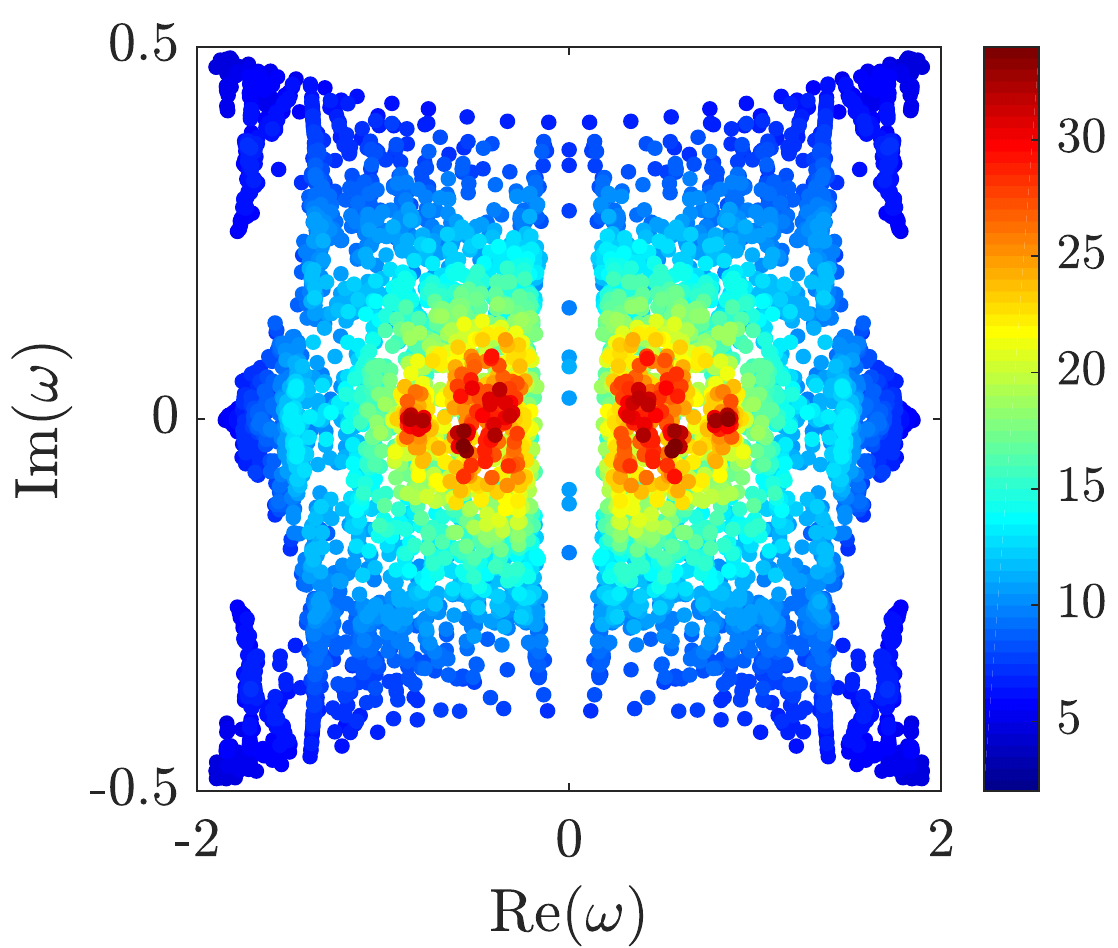}
	\caption{Localization length $\xi$ (colorbar) of the eigenstates of the non-Hermitian random pair-correlated lattice, as a function of the corresponding eigenvalue $\omega_{j}$ on the complex plane for $\alpha=0.5$. We can see that the localization length does not exceed the value of $\approx35$  ($N=2000$, five realizations of disorder are superimposed for visualization purposes).}
	\label{locl}
\end{figure}

 The next important issue we would like to address, is whether or not delocalization in this non-Hermitian model is possible. A direct and elegant way to see this is by considering a periodic lattice with a single dimer defect. In this case, one can analytically obtain an expression for the reflection probability-$|r|^2$ from the impurity \cite{dim1} [see Eq.~(\ref{refl}) in Appendix]. For the non-Hermitian dimer, the expression reads:
\begin{equation}
	|r|^2= \dfrac{\alpha^2(\cos^2{k}+\alpha^2)}{(1+2\alpha^2- 2\alpha \sin{k})^2-\cos^2{k}}.
\end{equation}
The above relation clearly shows that the equation $r=0$ cannot be satisfied for imaginary on-site energies. This indicates that, contrary to the Hermitian case, all the eigenstates are localized. In order to verify this statement, we calculate the localization length $\xi$ using the transfer matrix method \cite{Souk} and plot $\xi$ which corresponds to every eigenvalue on the complex plane, as is shown in Fig.~\ref{locl}. The localization length is defined as: 
\begin{equation}
\xi \equiv  -\lim\limits_{n \to \infty}\frac{n}{\ln|u_n|}. 
\end{equation}
In practice, if $N$ is much larger than $\xi$ (in our case, $N/\xi>50$), then one obtains a reliable value of $\xi$ as $n\to N$. We can clearly see that our assessment was correct, since $N/\xi$ does exceed the value of 50. Moreover, we find that $\xi$ remains finite for every value of $\alpha\neq0$. The fact that all eigenstates of the spectrum are localized, in contrast to the Hermitian case, is one of the most direct consequences of the non-Hermiticity on our system. As expected, the localization length is even smaller than the one shown in Fig.~\ref{locl} if the rectangular distribution, Eq.~(\ref{2a}), is used with the same standard deviation.

\section{NEW KIND OF TRANSPORT BY JUMPS}

\par Besides the interesting spectral properties of our model, phenomena associated with the spreading of an initially local excitation are the most intriguing and unexpected ones. These phenomena appear for all distributions of $\epsilon_{I,n}$ we have tried (see Appendices C and  D) and, hence,  we tend to believe that it is a universal feature associated with non-Hermitian open systems. In spite of the strong localization of the eigenstates, the initial local excitation seems to propagate not by diffusion (in this sense, Anderson’s conclusion for the absence of diffusion is still valid) but by a new kind of mechanism beyond Anderson, namely by apparently discontinuous jumps.  One way to systematically study the wave dynamics, is by examining the evolution pattern and the variance M$(z)$ of an initial single-channel, $n=n_{0}$ excitation as a function of $z$; $n_0$ was chosen to be at the center of the lattice: thus, 
$\psi_{n}(z=0)=\delta_{n,n_{0}}$. The variance is defined as follows:
\begin{equation}
\textnormal{M}(z)\equiv\sum_{n}(n-n_{0})^2|\psi_{n}(z)|^2~.
\end{equation}
\par
\begin{figure}[htb!]
	\centering
	\includegraphics[clip,width=1\linewidth]{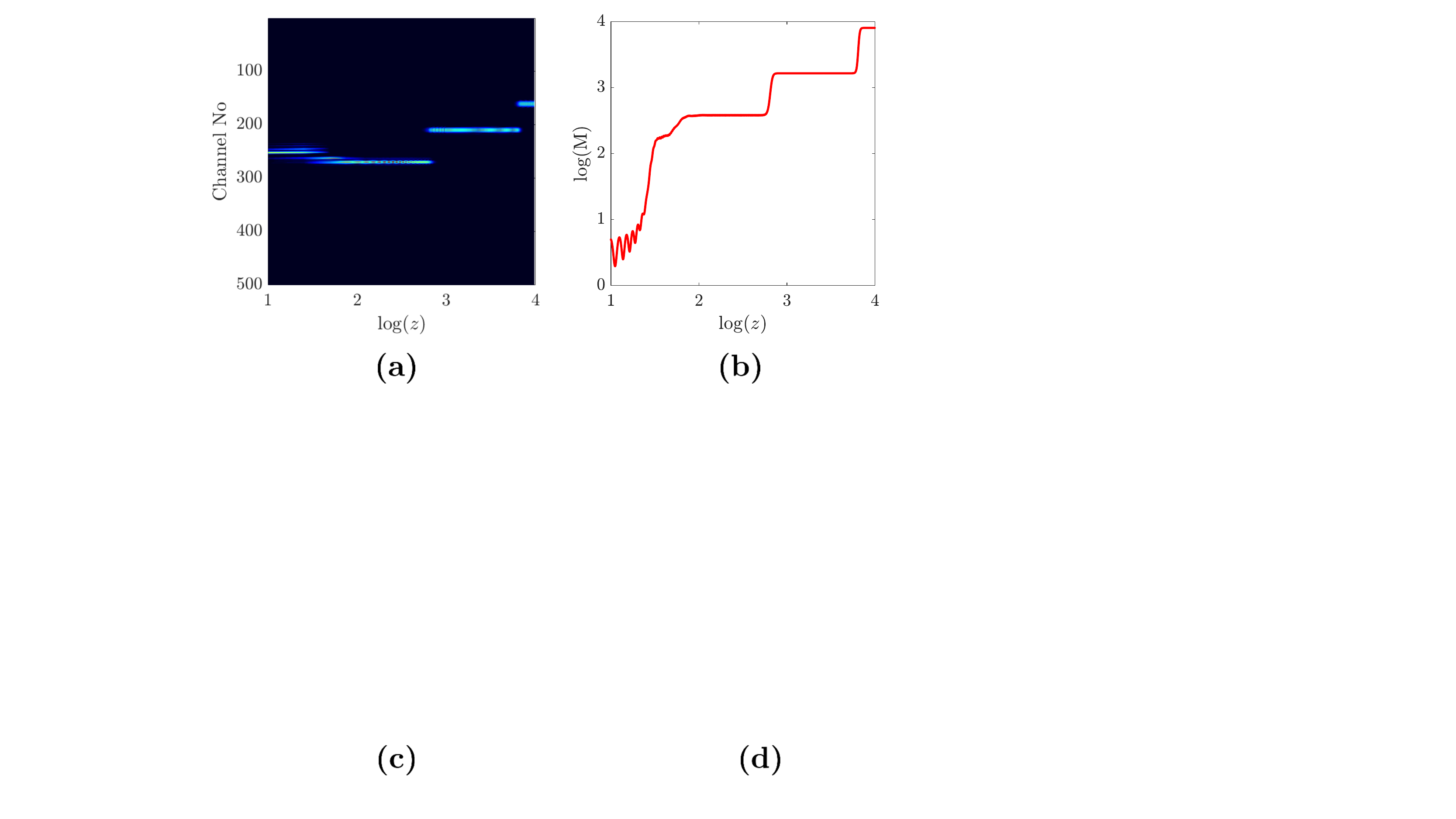}
	\caption{(a) Normalized field intensity $|\phi_{n}|^2$ as a function of the propagation distance's logarithm $\log{z}$. (b) Logarithm of the variance M$(z)$ as  a function of $\log{z}$ (here we have set $\alpha=1$ and $N=501$).}
	\label{dyn}
\end{figure}

 For $\alpha=0$ the lattice is periodic and $\textnormal{M}\sim z^2$, as expected, which indicates ballistic transport.  If we set $\alpha\neq0$, although all states become exponentially localized, then $\textnormal{M}(z)$ exhibits a very interesting and unexpected behavior. Our results are depicted in Fig.~\ref{dyn}. We note that we have plotted the power-normalized field amplitude: $|\phi_{n}|=\frac{|\psi_{n}|}{\sqrt{\mathcal{P}(z)}}$, since the field amplitude $|\psi_{n}|$ diverges exponentially as $z \to \infty$ due to the presence of gain. $\mathcal{P}(z) \equiv \sum_{n}|\psi_{n}|^2$ as to force energy conservation by this ``external" normalization.
\par Even though all the eigenstates are localized, the wave exhibits ``non-Hermitian jumps" between even distant sites and thus we obtain  energy spreading which seems to continue until the edges of the system are reached. This behavior is also captured by the plot of the variance over $z$ in Fig.~\ref{dyn}(b), for the same realization of disorder, which also exhibits a number of finite jumps. This behavior seems to occur in open systems with random complex $\epsilon_{n}$ and has not a Hermitian analog. In order to understand the physical mechanism behind this unexpected feature, we calculate the field at $z=z_{max}=10^4$ for the same realization of disorder as in Fig.~\ref{dyn}(a) and compare it with the field profile of the most gainy mode, namely the mode which corresponds to the eigenvalue with the largest value of the real part of $i\omega$ (see Appendix D).

\par In trying to physically interpret these intriguing results we have to stress two features of the eigenstates (left or right) which are unique in disordered random non-Hermitian systems: (a) their complex eigenvalues $\omega_j$, which lead to either infinite or zero amplitude as $z \to \infty$ depending on the sign of Im$(\omega_j)$, and (b) their nonorthogonality, which facilitates transfer of energy from channel to  channel, even between distant channels.  These two features seem to account for the jumpy spread of energy; indeed, as we have pointed out before, a gainy mode, i.e., one combining a large negative value of Im$(\omega_j)$ with a large overlap with the initial (or any intermediate) state is in a privileged situation (compared with a next neighboring channel) to be the recipient of an excitation. We  have confirmed this behavior by several numerical experiments in which we artificially introduced states with a large negative Im$(\omega_j)$ and located at selected sites. Always the jumps occur at these sites even when they were far away from the initial site. Can we conclude from these results that disordered random non-Hermitian systems with a short localization length allows a new type of transport by jumps to privileged eigenstates? Before we could provide a  positive answer to this question we have to consider an obvious possible reservation. The reason is that one has to be careful when referring to transport in non-Hermitian systems, because there are no intrinsic conservation laws. 
\par Notice, in this connection, that in our model at least two physical effects co-exist, one is the existence of eigenmodes of  quite different amplitude as a result of the range of values of the imaginary part of the eigenfrequencies inherent in an open non-Hermitian system, and the other is true transport of the energy associated with the initial excitation due to the coupling between adjacent channels and/or the nonorthogonality of the eigenstates. One may think at first that the observed jumps are an effect caused by gain. Thus to check this point we consider the dynamics of a non-Hermitian lattice $[-W_{R} \le \textnormal{Re}({\epsilon_n}) \le W_{R}; 0 \le \textnormal{Im}({\epsilon_n}) \le W_{I}: W_{R}=1,~W_{I}=2]$, with only loss (zero gain),  based on a rectangular random distribution. The intensity dynamics (power normalized to each step as before) is shown in the following Fig.~\textcolor{blue}{6}. 
\begin{figure}[htb!]
	\centering
	\includegraphics[clip,width=1\linewidth]{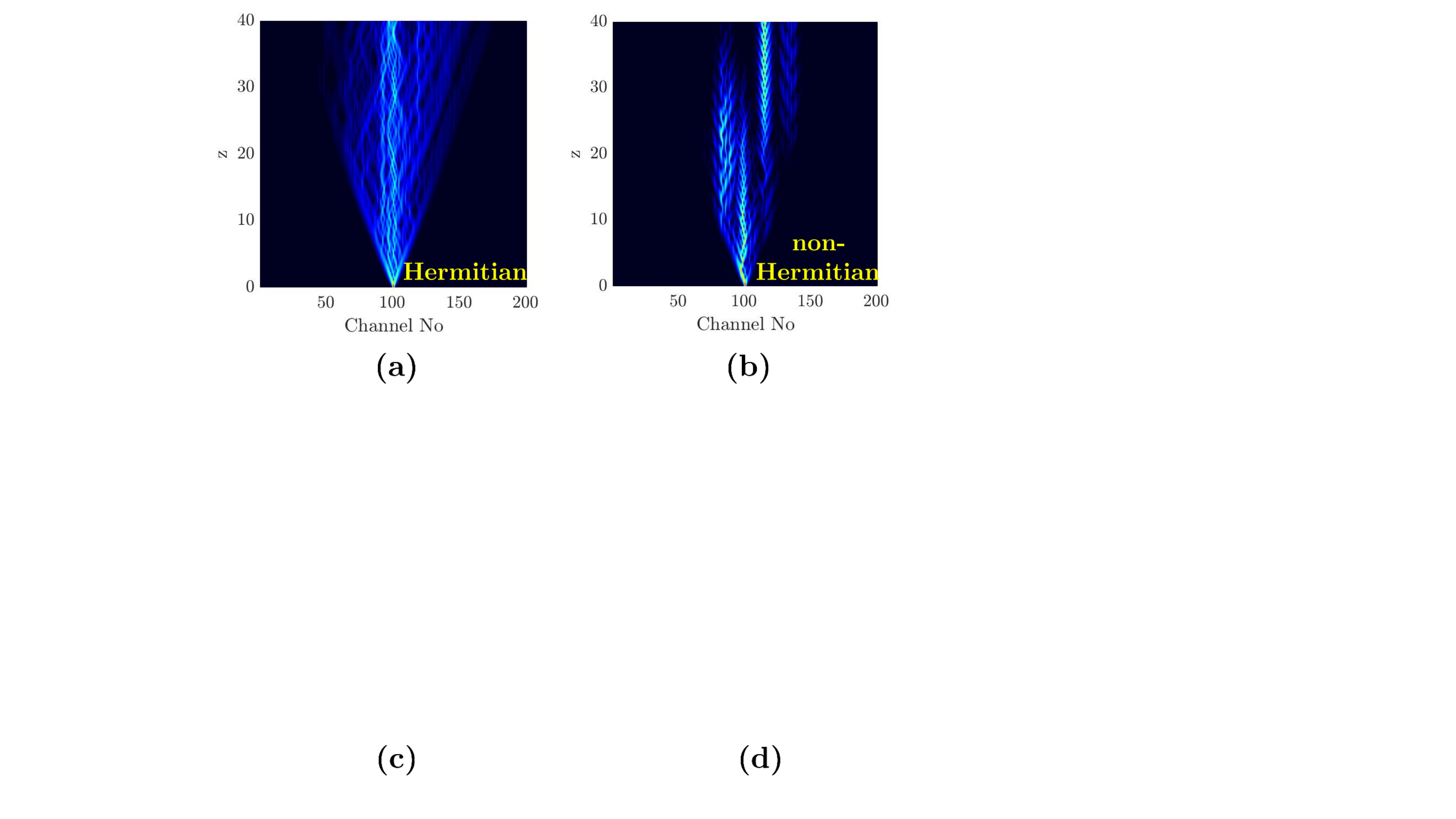}
	\caption{Evolution of a single-channel excitation ($N=201$) for (a) a disordered Hermitian lattice  ($W_R=1$), and (b)  a disordered non-Hermitian lattice with only loss ($W_R=1,~W_I=2$). The field is normalized is every step for visualization purposes.}
\label{mkr}
\end{figure}
As we can see the jumps are evident in the non-Hermitian lattice (right figure) in spite of the absence of any gain. Thus we can conclude that the external pumping of energy is not the reason for the jumps and, in the absence of any other way, we tend to conclude that what is observed is the actual transport of the initial excitation.  For comparison we show on the left of Fig.~\textcolor{blue}{6} the corresponding Hermitian lattice $(W_{R}=1,W_{I}=0)$ for the same realization but with no loss. We suspect that, in view of the elimination of amplification, the two physical mechanisms mentioned before are responsible for the physical origin of these jumps, namely the quite different amplitude of the eigenmodes and their nonorthogonality, two features which clearly distinguishes the non-Hermitian case from the Hermitian one. Note that this nonorthogonality can go to extreme in the sense that near an exceptional point (if any) the eigenvalues of the underlying system and the corresponding eigenvectors tend to simultaneously coalesce.
For the case of externally forced energy conservation and small $z$ no jumps were observed, although they appear for large enough $z$. Jumpy spread of the initial excitation was also found in the cases where normalization was replaced by gain saturation (as is quite usual experimentally) (see Appendix D). Thus, taking into account our own results and experimental \cite{Alex} and other work \cite{jumps1,jumps2} we suppose that the phenomenon of unusual jumpy transport can generally occur in Anderson-localized random non-Hermitian open systems, where the nonorthogonality of the eigenstates (left or right) and their quite different amplitude seem to play an important role.


\section{Conclusions}

\par In conclusion, we have systematically studied for the first time, the spectral and wave dynamic characteristics of non-Hermitian one-dimensional disordered lattices. Several probability distributions for the random imaginary matrix elements were considered. The  most interesting and surprising among our findings is the new kind of apparent transport in spite of strong localization of the eigenstates not by diffusion but through the sudden jumps even to distant sites; it seems that the nonorthogonality of the eigenmodes and their quite different amplitude play a significant role for what seems to be a new mechanism for transport in open non-Hermitian systems. For the case where the imaginary diagonal matrix element possess the binary pair-correlated disorder interesting spectral features were found:  for intermediate degree of randomness, the spectrum in the complex plane  has a fractallike intricate structure, while for higher values of the imaginary randomness parameter $\alpha$ many eigenvalues are concentrated in very small areas of the complex plane, a feature termed here ``eigenvalue condensation." 
We believe that this systematic study will open the way for the direct experimental realization of these phenomena in integrated photonic waveguide structures and for their further physical clarification.

\section{Acknowledgments}

\par We acknowledge support by the European Union's Horizon 2020 FETOPEN programme under project VISORSURF Grant No. 736876.

\appendix

\section{Biorthogonality relations and non-Hermitian eigenvalue problem}

\label{section:I}

In order to find the eigenmodes of the system, we substitute $\psi_{n,j}(z)=u_{n,j}e^{i\omega_j z}$ in the evolution equation [Eq.~(\ref{par})] and get the following eigenvalue problem:
\begin{equation}
c(u_{n+1,j}+u_{n-1,j})+\epsilon_{n,j}u_{n,j}=\omega_j u_{n,j},
\end{equation}
where $\omega_j$ is the complex eigenvalue of the $j^{th}$ eigenmode, j
with $j = 1, 2, ...,N$. In a more compact form the above eigenvalue problem can be expressed in terms of a symmetric tridiagonal matrix $D$ with zeros in the main diagonal, $c$ in the $\pm 1$ diagonals and the identity matrix $I$ (both matrices have dimension $N$ by $N$ ) by the following relation:
\begin{equation}
H_{ij}=D_{ij}+\epsilon_{i}I_{ij}
\end{equation}
Since $H$ is a non-Hermitian matrix, it is fully described by a set of biorthogonal right $\ket{u^R_j}$ and left $\ket{u^L_j}$ eigenmodes. In other words, we have the following right eigenvalue problem:
\begin{equation}
H \ket{u^R_j} =\omega_j \ket{u^R_j}
\end{equation}
and the corresponding left eigenvalue problem of the adjoint matrix:
\begin{equation}
H^{\dagger} \ket{u^L_j} =\omega^*_j \ket{u^L_j}
\end{equation}
The associated biorthogonality condition is:
\begin{equation}
\braket{u^L_j|u^R_i}=\delta_{ij}
\end{equation}
In general the right and left eigenvectors are different and, since the dynamics of the problem include both the right and the left set of eigenfunctions, one needs to study both of them. In our case though, the left and right eigenfunctions are complex conjugate pairs since $ H^{\dagger} =H^*$. This is a direct outcome of the last symmetry relation of the $H$ matrix.

\par Here we comment on the possibility of existence of a special type of spectral degeneracies, the exceptional points. In particular, exceptional points are non-Hermitian degeneracies that correspond to points in parameter space at which the eigenvalues of the underlying system and the corresponding eigenvectors simultaneously coalesce. In our system we did not find any exceptional points even in the case of parity-time symmetric disorder.

We comment also that the histogram  in Fig.~\ref{lvlsp}(a) seems to approach asymptotically the Wigner-Dyson distribution as the size of the system keeps increasing.

Finally, we point out that the majority of related non-Hermitian Anderson model studies, refer to the Hatano-Nelson Hamiltonian, which is non-Hermitian due to its asymmetric couplings (off diagonal elements of the matrix) and therefore is a nonreciprocal model. On the other hand, our Hamiltonian is reciprocal and its non-Hermiticity arises from the openness of the system, i.e., from the presence of dissipation and amplification (on the main diagonal of the matrix). That makes our model essentially different from the Hatano-Nelson Hamiltonian. Also our model has been experimentally realized in the context of guided wave and fiber loop optics \cite{Alex}, in contrast to the Hatano-Nelson model which is more difficult to realize in experiment. Beyond its theoretical importance our model is directly related to integrated photonic structures, and hence it is physically relevant.

\section{Hermitian binary disorder} 

\label{section:II}

\par The Hermitian one-dimensional model with binary disorder with or without short range correlations has been studied both theoretically and experimentally \cite{dim1,dim2,dim3,dim4}. Nevertheless, we redo here some of these  calculations in order to be able to compare  the resulting figures with new results associated with the non-Hermitian generalization.
\par Let us consider a waveguide array that exhibits a binary distribution  either $\epsilon_{a}$ or $\epsilon_{b}$ of its propagation constants (which here play the role of the on-site energies in the context of condensed matter physics), with the same probability $p=\frac{1}{2}$ and without any correlations:
\begin{equation}
\textnormal{Binary disorder :}~\epsilon_{n}=\begin{cases} \epsilon_{a}, \quad \textnormal{with} \quad \textnormal{p}_{1}=\frac{1}{2} \\
\epsilon_{b}, \quad \textnormal{with} \quad \textnormal{p}_{2}=\frac{1}{2}
\end{cases}
\end{equation}
where $\epsilon_{a,b}\in \mathbb{R}$ (Hermitian case). We also assume that the coupling coefficient between neighboring channels is constant and equal to $c$.
\par To begin with we examine the evolution pattern assuming a single-channel excitation in the middle of our lattice, namely: $\psi_{n}(z=0)=\delta_{n,n_{0}}$, where $n_{0}=\kappa+1$ (here we assume that the total site number is odd $N=2\kappa+1$.). More specifically, we are interested to consider the averaged variance of the intensity pattern as a function of the propagation distance $z$:
\begin{equation}
\textnormal{M}(z)=\langle\sum_{n}(n-n_{0})^2|\psi_{n}(z)|^2\rangle
\end{equation}
where $\langle..\rangle$ denotes averaging over many realizations of disorder.
\par
\begin{figure}[htb!]
\centering
\includegraphics[clip,width=1\linewidth]{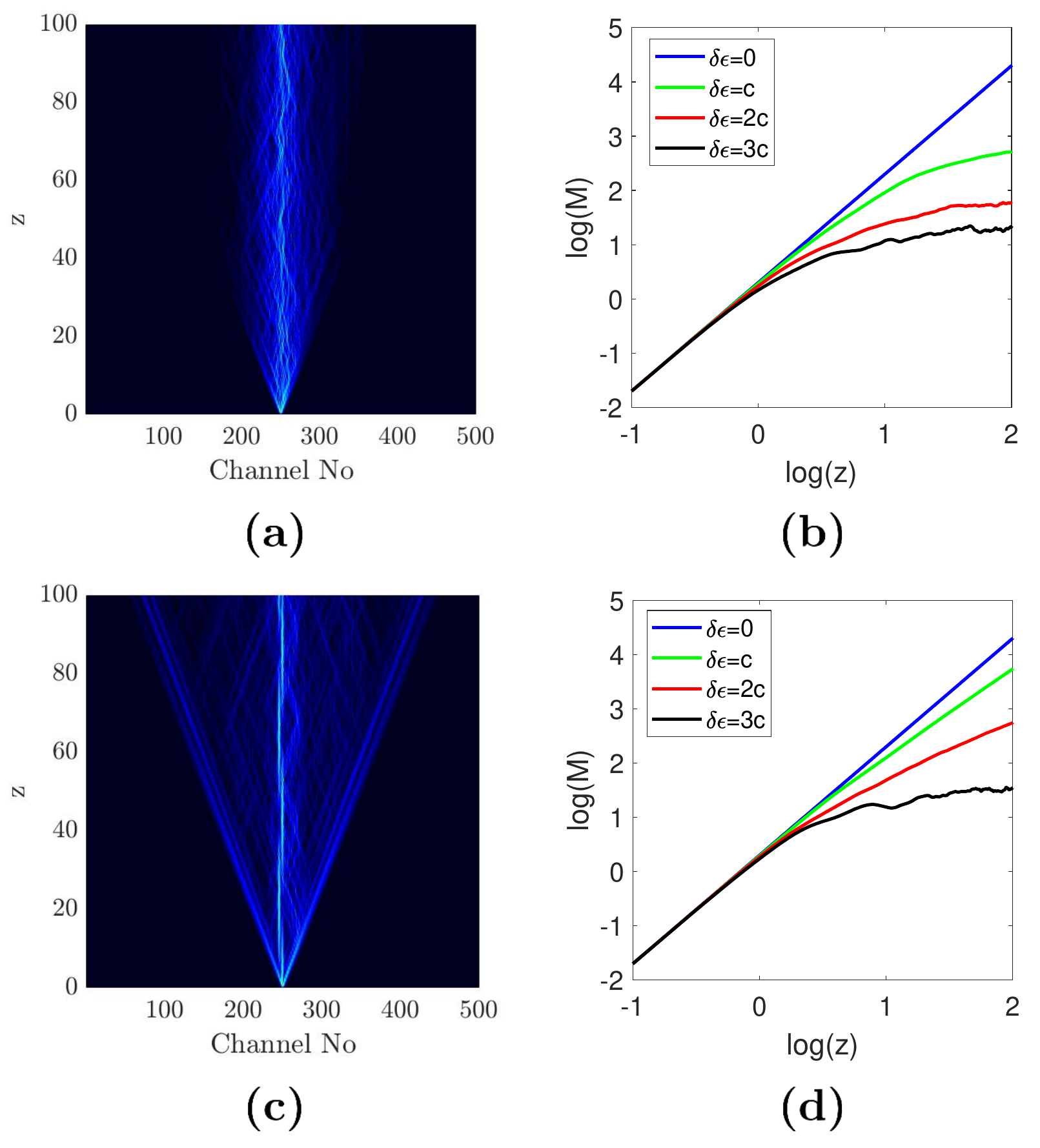}
\caption{(a) Field amplitude $|\psi_n|$ as a function of the propagation distance $z$, for the uncorrelated case with $\delta\epsilon=c$. (b) Logarithm of the averaged variance $M$ as a function of the logarithm of the propagation distance $z$ for the uncorrelated case and for different values of $\delta\epsilon/c$.  [c and d] Same as in [a and b] but for the random dimer lattice (see Ref.~\cite{dim4}). }
\label{fig1}
\end{figure}

 Our numerical calculations for this case are shown in Figs.~\ref{fig1}(a) and \ref{fig1}(b) and are in agreement with the corresponding experimental results \cite{dim4}. For $\epsilon_{a}=\epsilon_{b}\Rightarrow\delta\epsilon\equiv\epsilon_{a}-\epsilon_{b}=0$ the lattice is periodic and $\textnormal{M}\sim z^2$, which indicates ballistic transport. If we set $\epsilon_{a}\neq\epsilon_{b}$ though, then all the states become exponentially localized and $\textnormal{M}(z)$ saturates for large values of $z$; we get localization in this case since $\textnormal{M}\sim z^0$. The single-channel excitation remains localized near its initial position.
\par However, one gets completely different physical results if short-range order is introduced in this model. For that purpose, we now consider a ``dimer" waveguide array, where each dimer consists of two subsequent channels with the same propagation constant; this is a model originally introduced by Dunlap $et~al.$ \cite{dim1}:
\begin{equation}
\textnormal{Dimer array}:	\epsilon_{2n}=\epsilon_{2n+1}=\begin{cases} \epsilon_{a}, \quad \textnormal{with} \quad p_{1}=\frac{1}{2} \\
\epsilon_{b}, \quad \textnormal{with} \quad p_{2}=\frac{1}{2}
\end{cases}~.
\end{equation}

\par Repeating the same calculations as in Ref.~\cite{dim1}, we can see that now, $M\sim z^{\gamma}$, in all the cases, where $\gamma\simeq$ 2, $\frac{3}{2}$, 1 and 0, which correspond to ballistic, superdiffusive, diffusive and localized motion, accordingly (computed numbers are: 1.99, 1.56, 0.98 and 0.15) for $\dfrac{\delta\epsilon}{c}=$ 0, 1, 2 and 3 respectively. These results indicate that the spectrum now possesses delocalized eigenvectors. Indeed, one can prove that eigenstates with eigenvalue $\omega=\epsilon_{a,b}$ are extended, as long as $|\delta\epsilon|\leq2c$. All the relevant results are presented in Fig.~\ref{fig1} and are again in perfect agreement with the experimental ones \cite{dim4}.
\par A direct way to obtain a physical insight of these results is to consider a periodic lattice with a single dimer defect. One can show that the reflection probability from the impurity is given by the following expression \cite{dim1}:
\begin{equation}
|r|^2=\dfrac{\delta\epsilon^2[\delta\epsilon+2c\cos(k)]^2}{\delta\epsilon^2[\delta\epsilon+2c\cos(k)]^2+4c^4\sin^2(k)}
\label{refl}
\end{equation}
where $k\equiv\cos^{-1}(\omega/2c)$ is the Bloch wave number. From the expression above one can easily show that waves with $\omega=\epsilon_{a,b}$ are perfectly transmitted ($r=0$), provided that $|\delta\epsilon|\leq2c$. Furthermore, it was found that the total number of states with localization length greater than the system's size is of measure $\sqrt{N}$ \cite{dim1}. Thus, in Fig.~\ref{fig1}(c), the two propagating peaks correspond to these $\sim\sqrt{N}$ delocalized states, leading to transport, while the central peak indicating no propagation is associated with the vast majority of localized states. 
\par It is interesting to consider the criterion for localization originally proposed by Anderson \cite{loc3}; it was the asymptotic behavior of the amplitude of the wave function around its initial site, in the sense that absence of diffusion is associated with the limit: $\lim_{z\to\infty}|\psi_{n_{0}}(z)|$ being nonzero. Thus, it is reasonable to examine the probability $P(z)$ for the wave to be located in its initial position as a function of the propagation distance for the two cases of disorder discussed here. 

\begin{figure}[htb!]
\centering
\includegraphics[clip,width=1\linewidth]{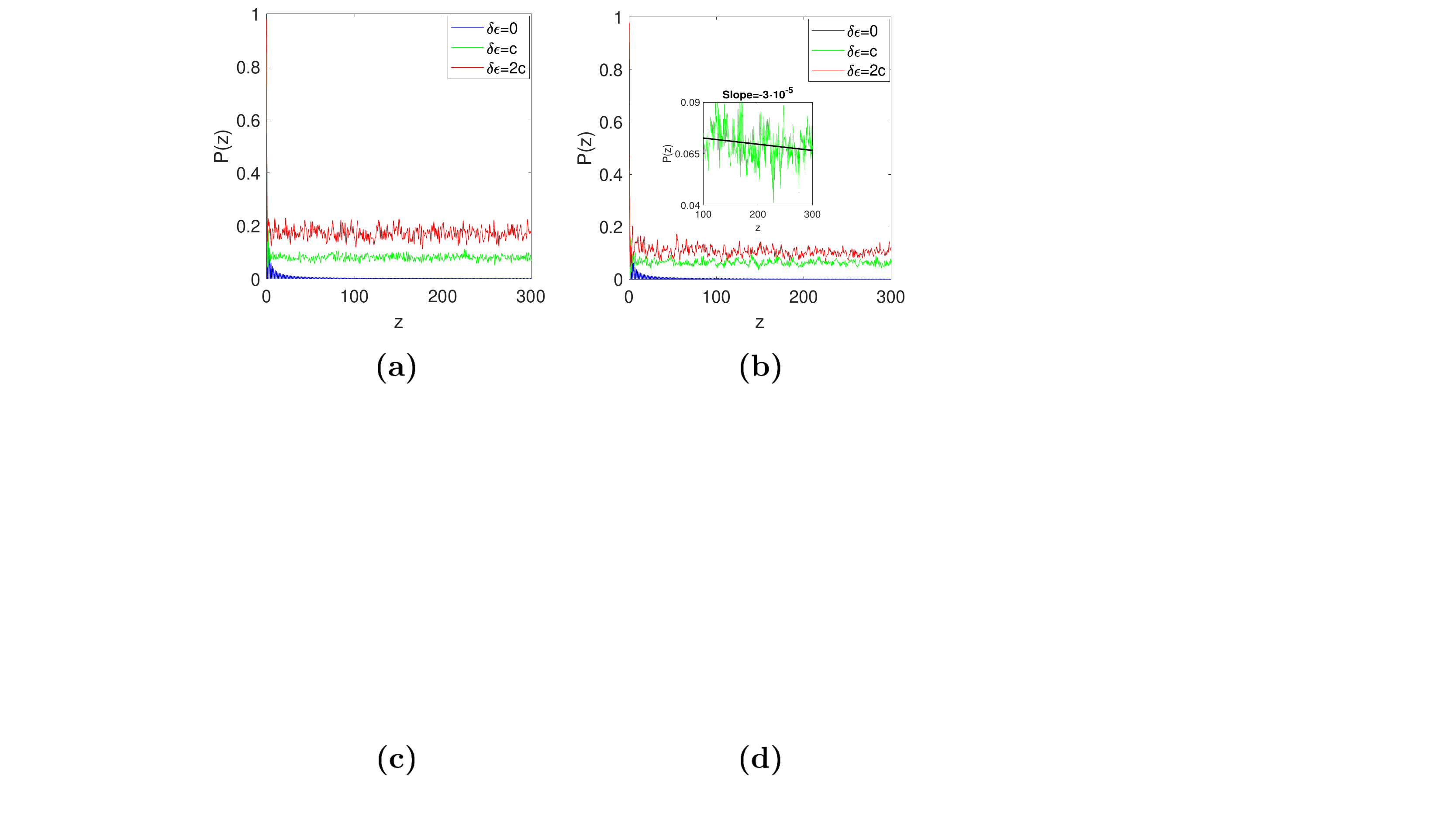}
\caption{Probability $P(z)$ for the wave to being found in its initial position as a function of the propagation distance $z$ for (a) the uncorrelated binary disorder and (b) the dimer case and for different values of $\delta\epsilon/c$. An averaging over 50 realizations of disorder has been performed for each plot. Inset: A zoom in the plot of $P$ vs. $z$ for $\delta\epsilon=c$ (green line) and the least- squares fitting of the curve (black line). The slope of the line is shown in the title of the graph. }
\label{b1}
\end{figure}
\par In Fig.~\ref{b1} we show plots of $P(z)$ for the uncorrelated binary [Fig.~\ref{b1}(a)] and the dimer array [Fig.~\ref{b1}(b)], for $\delta\epsilon/c=0,1$ and $2$. The difference between the two graphs is small but crucial. While in Fig.~\ref{b1}(a), and for $\delta\epsilon\ne0$ $P(z)$, fluctuates around a specific, constant value, in Fig.~\ref{b1}(b) $P(z)$ slowly drops as $z$ increases, with a slope of $\sim10^{-5}$, which is actually the value of the localization length for $\omega$ near $\epsilon_{1,2}$. This is shown clearly in the inset of Fig.~\ref{b1}(b), where a least square fit (black line) is also plotted with $P(z)$. This statement is in agreement with Anderson criterion of localization, as it should be. However, due to the many fluctuations and the very small value of the linear fitting's slope, $P(z)$ is not a convenient numerical criterion for localization in this case.

\section{Dynamics in Non-Hermitian binary random lattices}

\label{section:III}

\par In this section we discuss further new results regarding the phenomenon of jumpy transport\cite{Alex} in our non-Hermitian binary pair-correlated disordered model.
\par
\begin{figure}[htb!]
\centering
\includegraphics[clip,width=1\linewidth]{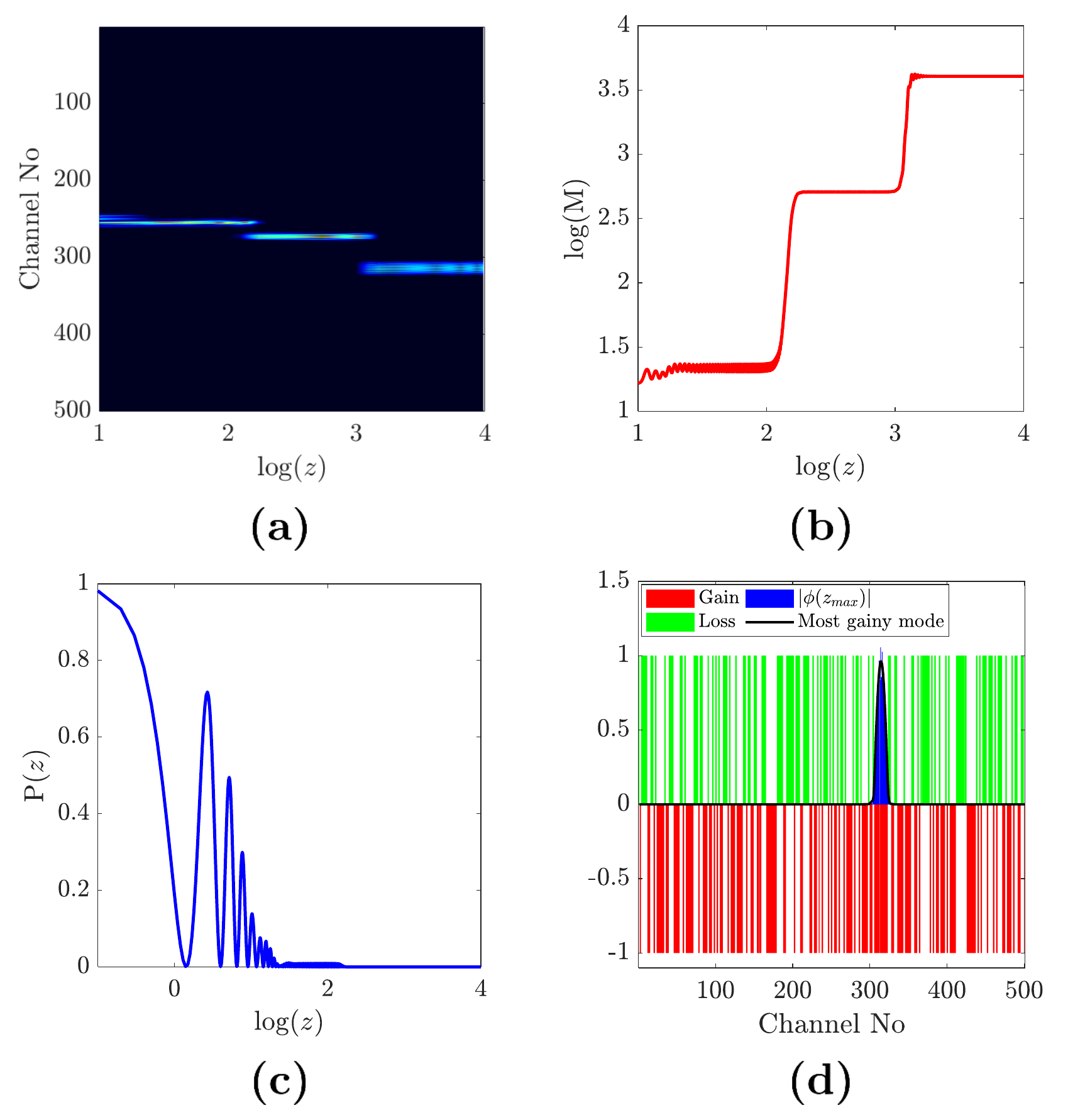}
\caption{(a) Normalized field amplitude $|\phi_n|$ as a function of the logarithm of the propagation distance $z$, under single-channel excitation and a particular realization of the random lattice with $\alpha=1$ (see Eq.~(\ref{2b}) in main text). (b) Logarithm of the averaged variance $M$ as a function of the logarithm of the propagation distance $z$. (c) Probability $P$ for the wave to return to its initial position as a function of the propagation distance $z$. (d) Normalized field profile $|\psi_n|$ near the end of the lattice $z=z_{max}$ (blue bars). The (normalized) field profile of the most gainy eigenstate is also plotted here for comparison (black line). The gain and loss distributions (imaginary part of the potential) are depicted with red, green color, respectively. In all the above results the number of channels was N=501.}
\label{fig2}
\end{figure}

 Let us consider the wave evolution of a single-channel excitation (in the middle of the lattice), as a function of the propagation distance $z$. Since the spectrum contains eigenvalues that correspond to amplification, we  amplitude normalize the field in every step in order to obtain a physically meaningful diffraction pattern. The result is depicted for a particular realization in Fig.~\ref{fig2}(a). We can clearly see a finite number of jumps in the transverse direction. This dynamical behavior is also reflected in the  step-like discontinuities of the variance $M$ as is shown in the plot of the logarithm of the variance as a function of $\log(z)$, in Fig.~\ref{fig2}(b), as well as in the abrupt drop in the amplitude of the initial site $n_{0}$, in Fig.~\ref{fig2}(c).
\par In order to highlight the underlying physical mechanism of these transverse jumps, we calculate the eigenstate that corresponds to the most gainy eigenvalue (black line in Fig.~\ref{fig2}(d)). The single-channel excitation at $z=0$ excites many localized eigenmodes that have complex eigenvalues. The superposition of these modes generates a complex evolution pattern as a result of the interference of these nonorthogonal eigenstates. No matter how small is the amplitude of the projection coefficient that corresponds to the most gainy eigenvalue, for long propagation distances it will always dominate over the other modes. Therefore, the location of the final jump is solely determined by the most gainy eigenstate. This physical explanation of the jumps is also supported by direct numerical simulations. In Fig.~\ref{fig2}(d) we can clearly see that the field  near the end of the lattice ($z=z_{max}$) is almost the same with the  field profile of the most gainy eigenstate, as we expected. 

\section{Comments on the dynamics in Non-Hermitian Anderson lattices}

\label{section:IV}

\begin{figure}[htb!]
	\centering
	\includegraphics[clip,width=1.0\linewidth]{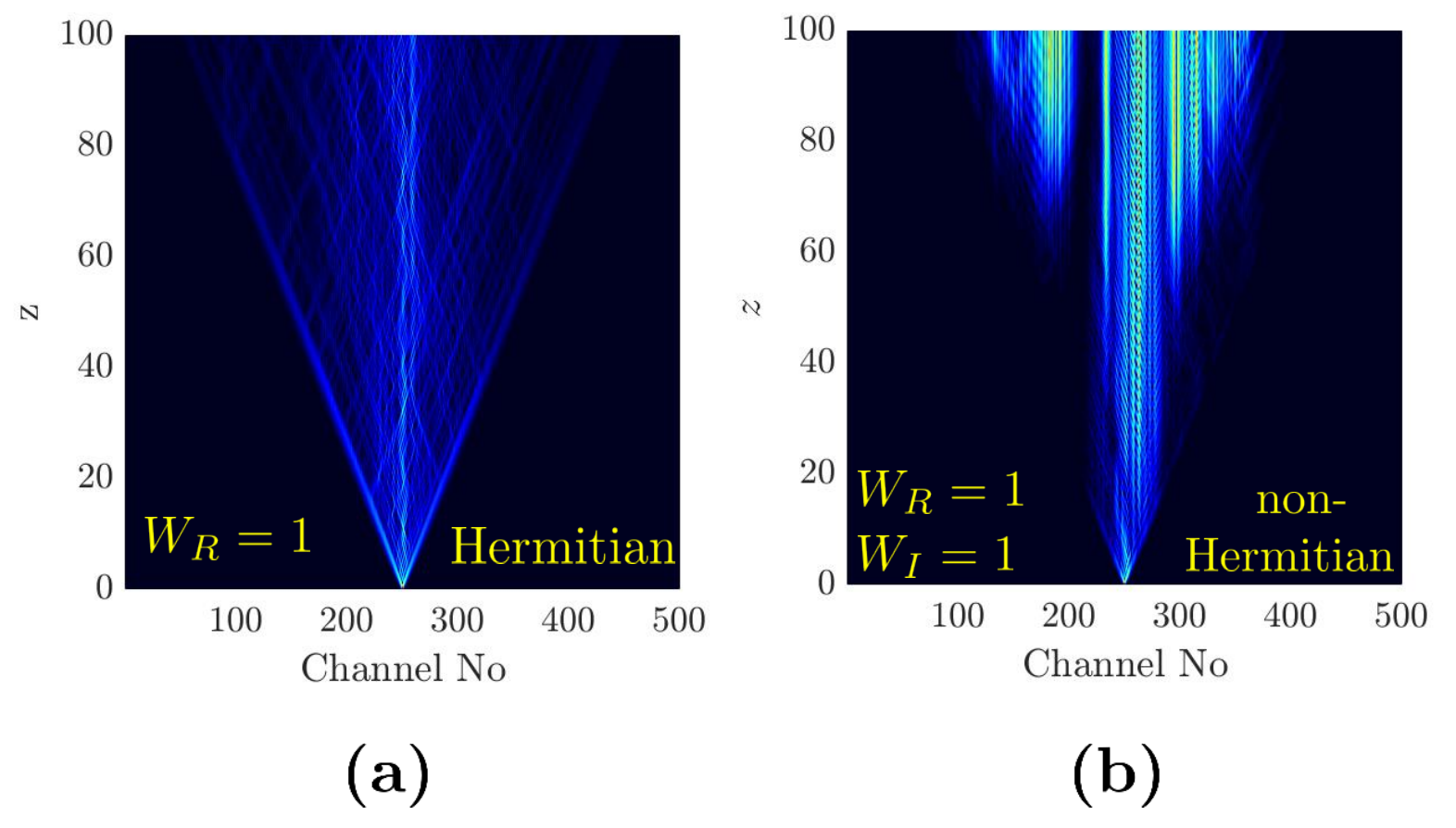}
	\caption{Field amplitude $|\psi_n|$ (not normalized) as a function of $z$ for (a) Hermitian disordered lattice ($W_R=1$) and (b) non-Hermitian disordered lattice with saturable gain of the form $i \textnormal{Im}(\epsilon_n)/(1+\mu|\psi_n|^2 )$ ($\mu=5, W_R=W_I=1 $). The jumps are evident here as well.}
	\label{10}
\end{figure}

In this section we further discuss several aspects regarding the dynamics in non-Hermitian disordered lattices with uncorrelated disorder in both the real and imaginary parts. Indeed one has to be careful when referring to transport for non-Hermitian systems, because there are no conservation laws. In the main text we discuss what happens if the random non-Hermitian system possesses only loss and no gain at all. We found that jumps are present in this case as well, thus excluding the external supply of energy as an explanation and apparently leaving the true transport by jumps as the real new mechanism for propagation in these systems.  We examined also  what happens if we do not consider any normalization, for both weak and strong disorder. For weak disorder we can demonstrate clearly the effect of gain/loss that leads to unique asymmetric transport dynamics. Notice this picture is true even for the realizations that the total power decreases with the not so large propagation distance. The asymmetry of the pattern seems to be the direct outcome of eigenstate nonorthogonality; in such short propagation distances we do not see jumps. The  features of the non-Hermitian transport though, are clear without performing any normalization. For strong disorder we examine what happens where the gain is of saturable type of the form $i \textnormal{Im}(\epsilon_n)/(1+\mu |\psi_n|^2 )$; indeed for strong disorder we consider the case of $W_{R}=1, -W_{I} \le \textnormal{Im}(\epsilon_n) \le W_{I}, W_{I}=1, \mu=5$ (Fig.~\ref{10}). The jumps are evident here as well. Note that here we do not normalize the field, and the saturation of the gain limits the amplification effect of eigenmodes.
\par The results of our findings can be summarized in three basic conclusions: (a) Field normalization in every propagation step is a way to visualize the center of mass displacement and is also experimentally possible in the fiber loop set-up. (b) Jumps occur also in purely dissipative lattices (with zero gain) and in forcibly energy conserving normalization, apparently as a result of eigenstates nonorthogonality and their quite different amplitude. (c) Inclusion of saturable nonlinear gain also allows for similar abrupt jumps in a random non-Hermitian lattice and can be directly observed without the need of any field normalization. 
\par In conclusion, transport in non-Hermitian lattices is a difficult and fundamental topic of wave physics and the existence of sudden jumps demonstrates its complexity.  There are still various open questions regarding the nature of wavepacket evolution in such open systems.

\bibliographystyle{longbibliography}

\begin{thebibliography}{100}



\bibitem{loc3}P. W. Anderson, Phys. Rev. {\bf109}, 1492 (1958).
\bibitem{loc1}E. Akkermans and G. Montambaux, {\it Mesoscopic Physics of Electrons and Photons} (Cambridge University Press, Cam- bridge, 2007).
\bibitem{loc11} I. M. Lifshitz, S. A. Gredeskul, and L. A. Pastur, Introduction to the theory of disordered systems (WileyInterscience, 1988).
\bibitem{loc12}N. F. Mott and E. A. Davis, Electronic processes in non-crystalline materials (Oxford university press, 2012).
\bibitem{loc13} P. W. Anderson, Philos. Mag. B, {\bf52}, 505-509 (1985).
\bibitem{loc14} D. S. Wiersma, P. Bartolini, A. Lagendijk and R. Righini, Nature, {\bf390}, 671-673 (1997).
\bibitem{loc4} A. Lagendijk, B. Van Tiggelen, and D. S. Wiersma, Phys.Today, {\bf62}, 24 (2009).
\bibitem{loc5} T. Schwartz, G. Bartal, S. Fishman, and M. Segev, Nature {\bf446}, 52 (2007).
\bibitem{loc6}Y. Lahini, A. Avidan, F. Pozzi, M. Sorel, R. Morandotti, D. N. Christodoulides, and Y. Silberberg, Phys. Rev. Lett. {\bf100}, 013906 (2008).
\bibitem{rl1}D. S.  Wiersma, Nat. Phys., {\bf4}, 359-367 (2008).
\bibitem{rl2}H. Cao, et. al., Phys. Rev. Let., 82, 2278 (1999).
\bibitem{rl3} C. Vanneste and P. Sebbah, Phys. Rev. Let., {\bf87}, 183903 (2001).



\bibitem{Bender1} C.~M.~Bender and S.~Boettcher, Phys. Rev. Lett. {\bf 80}, 5243 (1998).
\bibitem{Bender2} C.~M.~Bender, S.~Boettcher, and P.~N.~Meisinger, J. Math. Phys. {\bf 40}, 2201 (1999).
\bibitem{Bender3} C.~M.~Bender, D.~C.~Brody, and H.~F.~Jones, Phys. Rev. Lett. {\bf 89}, 270401 (2002).
\bibitem{PT1} K.~G.~Makris, R.~El-Ganainy, D.~N.~Christodoulides, and Z.~H.~Musslimani, Phys. Rev. Lett. {\bf 100}, 103904 (2008).
\bibitem{PT2} R.~El-Ganainy, K.~G.~Makris, D.~N.~Christodoulides, and Z.~H.~Musslimani, Opt. Lett. {\bf 32}, 2632 (2007).
\bibitem{PT3} Z.~H.~Musslimani, K.~G.~Makris, R.~El-Ganainy, D.~N.~Christodoulides, Phys. Rev. Lett. {\bf 100}, 030402 (2008).
\bibitem{PT4} A. Guo, G. J. Salamo, D. Duchesne, R. Morandotti, M.
Volatier-Ravat, V. Aimez, G. A. Siviloglou, and D. N. Christodoulides, Phys. Rev. Lett. {\bf 103}, 093902 (2009).
\bibitem{PT5} C. E. R\"uter, K. G. Makris, R. El-Ganainy, D. N. Christodoulides, M. Segev, and D. Kip, Nat. Phys. {\bf 6}, 192 (2010).
\bibitem{EP1} N. Moiseyev, {\it Non-Hermitian Quantum Mechanics} (Cambridge, New York, 2011).
\bibitem{EP2} M. V. Berry, Czechoslovak J. Phys. {\bf 54}, 1039 (2004).
\bibitem{EP3} W. D. Heiss, J. Phys. A: Math. Gen. {\bf 37}, 2455 (2004).
\bibitem{EP4} J. Wiersig, S.-W. Kim, and M. Hentschel, Phys. Rev. A {\bf 78}, 053809 (2008).
\bibitem{EP5} S.-B. Lee {\it et al.}, Phys. Rev. Lett. {\bf 103}, 134101 (2009).


\bibitem{review1} V. V. Konotop, J. Yang, and D. A. Zezyulin, Rev. Mod. Phys. {\bf88}, 035002 (2016).
\bibitem{review2} L. Feng, R. El-Ganainy, and L. Ge, Nat. Phot.  {\bf 11}, 752 (2017).
\bibitem{Pile} D.F. Pile, and D. N. Christodoulides, Nat. Phot. {\bf 11}, 742 (2017).
\bibitem{Gbur} G. Gbur, and K.G. Makris, Photonics Research  {\bf 6}, PTS1 (2018).
\bibitem{review3} R. El-Ganainy, K.G. Makris, M. Khajavikhan, Z.H. Musslimani, S. Rotter, and D. N. Christodoulides, Nat. Phys.  {\bf 14}, 11 (2018).
\bibitem{Longhi} S. Longhi, EPL {\bf120}, 64001 (2018).
\bibitem{review4} S.K. \"Ozdemir, S. Rotter, F. Nori, and L. Yang, Nat. mater.  {\bf 18}, 783 (2019).
\bibitem{review5} M.A. Miri and A. Alu, Science {\bf 363}, eaar7709 (2019). 

\bibitem{PT6} Y.~D.~Chong, L.~Ge, and A.~D.~Stone, Phys. Rev. Lett. {\bf 106}, 093902 (2011).
\bibitem{PT7} L.~Ge, Y.~D.~Chong, and A. D. Stone, Phys. Rev. A {\bf 85}, 023802 (2012).
\bibitem{PT8} P.~Ambichl, K.~G.~Makris, L.~Ge, Y.~D.~Chong, A.~D.~Stone, and S.~Rotter, Phys. Rev. X {\bf 3}, 041030 (2013).
\bibitem{PT9} A. Regensburger, C. Bersch, M.-A. Miri, G. Onishchukov, D. N. Christodoulides, and U. Peschel, Nature {\bf 488}, 167 (2012).
\bibitem{PT10} L. Feng,	Y.-L. Xu,	W. S. Fegadolli,	 M.-H. Lu,	J. E. B. Oliveira, V. R. Almeida, Y.-F. Chen, and A. Scherer, Nat. Mater. {\bf 12}, 108 (2013).
\bibitem{PT11} B. Peng, S. K. Ozdemir, F. Lei, F. Monifi, M. Gianfreda, G. L. Long, S. Fan, F. Nori, C. M. Bender, and L. Yang, Nat. Phys. {\bf 10}, 394 (2014).
\bibitem{PT12} L. Feng, Z. Jing Wong, R.-M. Ma, Y. Wang, and X. Zhang, Science {\bf 346}, 972 (2014).
\bibitem{PT13} H. Hodaei, M.-A. Miri, M. Heinrich, D. N. Christodoulides, and M. Khajavikhan, Science {\bf 346},
975 (2014).
\bibitem{PT14} B. Peng, S. K. \"Ozdemir, S. Rotter, H. Yilmaz, M. Liertzer, F. Monifi, C. M. Bender, F. Nori, and L. Yang Science {\bf 346}, 328 (2014).
\bibitem{PT15} S. Assawaworrarit, X. Yu, and S. Fan, Nature {\bf546}, 387 (2017).
\bibitem{PT16} J. Zhang, B. Peng, S.K. \"Ozdemir, K. Pichler, D.O. Krimer, G. Zhao, F. Nori, Y. Liu, S. Rotter, and L. Yang, Nat. Phot. {\bf 12}, 479 (2018).


\bibitem{lin_unidirectional_2011} Z. Lin, H. Ramezani, T. Eichelkraut, T. Kottos, H. Cao, and D.N. Christodoulides, Phys. Rev. Lett. {\bf106}, 213901 (2011).
\bibitem{horsley_spatial_2015} S. A. R. Horsley, M. Artoni, and G. C. La Rocca, Nat. Phot. {\bf9}, 436 (2015).
\bibitem{konotop_families_2014} V. V. Konotop and D. A. Zezyulin, Opt. Lett. {\bf39}, 5535 (2014).
\bibitem{zhu_one-way_2013} X. Zhu, L. Feng, P. Zhang, X. Yin, and X. Zhang, Opt. Lett. {\bf38}, 2821 (2013).

\bibitem{Andreas}A. F. Tzortzakakis, K. G. Makris, and E. N. Economou Phys. Rev. B {\bf101}, 014202 (2020).
\bibitem{nhloc1} Y. Huang and B. I. Shklovskii, Phys. Rev. B, {\bf101}, 014204 (2020).
\bibitem{nhloc2}P. O. Sukhachov and A. V. Balatsky, Phys. Rev. Res., {\bf2}, 013325 (2020).
\bibitem{nhloc3}Y. Liu, X. P. Jiang, J. Cao, J. and S. Chen, Phys. Rev. B, {\bf101}, 174205 (2020).
\bibitem{nhloc4}K. Kawabata, K. and S. Ryu, arXiv preprint arXiv:2005.00604 (2020).
\bibitem{nhloc5}H. Liu, Z. Su, Z. Q. Zhang and H. Jiang, Chin. Phys. B {\bf{29}}, 050502 (2020).
\bibitem{Alex} S. Weidemann, M. Kremer, S. Longhi and A. Szameit, in Conference on Lasers and Electro-Optics, OSA Technical Digest (Optical Society of America, 2020), paper FTu3A.3.
\bibitem{nhloc6} J. Feinberg and A. Zee, Phys. Rev. E {\bf59}, 6433 (1999).
\bibitem{nhloc7} N. Hatano and D. R. Nelson, Phys. Rev. B {\bf58}, 8384 (1998).
\bibitem{nhloc8} L G Molinari,, J. Phys. A Math. Theor., {\bf42}, 265204 (2009)
\bibitem{nhloc9}  A Amir, N. Hatano, and D. R. Nelson, Physical Review E {\bf93}, 042310 (2016).



\bibitem{dim1} D. H. Dunlap, H. L. Wu and P. W. Phillips, Phys. Rev. Let., {\bf65}, 88 (1990).
\bibitem{dim2} P. Phillips and H. L. Wu, Science, {\bf252}, 1805-1812 (1991).
\bibitem{dim3} V. Bellani, E. Diez, R. Hey, L. Toni, L. Tarricone, G. B. Parravicini, and R. Gómez-Alcalá, Phys. Rev. Let., {\bf82}, 2159 (1999).
\bibitem{dim4} U. Naether, S. Stützer, R. A. Vicencio, M. I. Molina, A. Tünnermann, S. Nolte, and A. Szameit, New Jour. of Phys., {\bf15}, 013045 (2013).
\bibitem{harp} P. G. Harper, Proc. Phys. Soc. {\textbf68}, 874 (1955).
\bibitem{Souk} P. Markos and C. M. Soukoulis, Wave propagation: from electrons to photonic crystals and left-handed materials (Princeton University Press, Princeton, NJ, 2008).
\bibitem{jumps1}I. I. Yusipov, T. V. Laptyeva and M. V. Ivanchenko, Phys. Rev. B, {\bf97}, 020301 (2018).
\bibitem{jumps2}M. Balasubrahmaniyam, S. Mondal, S. and S. Mujumdar, Phys. Rev. Let., {\bf124}, 123901 (2020).

\end{thebibliography}

\end{document}